\input harvmac.tex

\noblackbox
\lref\oldjd{J. Distler, ``Resurrecting (2,0) Compactifications,''
Phys. Lett. {\bf B188} (1987) 431.} 
\lref\ahw{I. Affleck, J. Harvey, and E. Witten, ``Instantons and
(Super)Symmetry Breaking in (2+1) Dimensions,'' Nucl. Phys.
{\bf B206} (1982) 413.}
\lref\mayr{P. Mayr, ``Mirror Symmetry, N=1 Superpotentials, and
Tensionless Strings on Calabi-Yau Fourfolds,'' hep-th/9610162.}
\lref\aspg{P. Aspinwall and M. Gross, ``The SO(32) Heterotic String
on a K3 Surface,'' Phys. Lett. {\bf B387} (1996) 735, hep-th/9605131.}
\lref\oldcand{P. Green and T. Hubsch, ``Phase Transitions Among
(Many of) Calabi-Yau Compactifications,'' Phys. Rev. Lett. {\bf 61} (1988)
1163\semi
P. Candelas, P. Green, and T. Hubsch, ``Finite Distances Between
Distinct Calabi-Yau Vacua,'' Phys. Rev. Lett. {\bf 62} (1989) 1956.}
\lref\andy{A. Strominger, ``Massless Black Holes and
Conifolds in String Theory,'' Nucl. Phys. {\bf B451} (1995) 96, 
hep-th/9504090\semi
B. Greene, D. Morrison, and A. Strominger, ``Black Hole Condensation
and the Unification of String Vacua,'' Nucl. Phys. {\bf B451} (1995) 109,
hep-th/9504145\semi
T. Chiang, B. Greene, M. Gross, and Y. Kanter, ``Black Hole Condensation
and the Web of Calabi-Yau Manifolds,'' hep-th/9511204\semi
A. Avram, P. Candelas, D. Jancic, and M. Mandelberg, ``On the
Connectedness of Moduli Spaces of Calabi-Yau Manifolds,'' Nucl. Phys.
{\bf B465} (1996) 458, hep-th/9511230.} 
\lref\sixd{O. Ganor and A. Hanany, ``Small $E_8$ Instantons and
Tensionless Noncritical Strings,'' Nucl. Phys. {\bf B474} (1996) 122,
hep-th/9602120\semi
N. Seiberg and E. Witten, ``Comments on String Dynamics in Six
Dimensions,'' Nucl. Phys. {\bf B471} (1996) 121, hep-th/9603003.}
\lref\vw{C. Vafa and E. Witten, ``Dual String Pairs with N=1 and N=2
Supersymmetry in Four Dimensions,'' hep-th/9507050.} 
\lref\edsusy{E. Witten, ``Strong Coupling Expansion of Calabi-Yau
Compactification,'' Nucl. Phys. {\bf B471} (1996) 135, hep-th/9602070.}
\lref\kss{S. Kachru, N. Seiberg, and E. Silverstein, ``SUSY Gauge
Dynamics and Singularities of 4d N=1 String Vacua,'' {Nucl. Phys.}
{\bf B480} (1996) 170, hep-th/9605036.}
\lref\edphases{E. Witten, ``Phases of N=2 Theories in Two Dimensions,''
Nucl. Phys. {\bf B403} (1993) 159, hep-th/9301042.}
\lref\silvwitt{E. Silverstein and E. Witten, ``Criteria for Conformal
Invariance of (0,2) Models,'' {Nucl. Phys.} {\bf B444} (1995) 161, 
hep-th/9503212.}
\lref\VafaF{C. Vafa, ``Evidence for F theory,'' Nucl. Phys. {\bf B469}
(1996) 403, hep-th/9602022.}
\lref\FII{D. Morrison and C. Vafa, ``Compactifications of
F-theory on Calabi-Yau Threefolds II,'' Nucl. Phys. {\bf B476} (1996) 437,
hep-th/9603161.}
\lref\nonchiral{
I. Brunner, M. Lynker, and R. Schimmrigk, ``Unification of M Theory
and
F Theory Calabi-Yau Fourfold Vacua,'' hep-th/9610195\semi
A. Klemm, B. Lian, S.S. Roan, and S.T. Yau, 
``Calabi-Yau Fourfolds for M Theory and F Theory Compactifications,''
hep-th/9701023\semi
K. Mohri, ``F Theory Vacua in Four-Dimensions and Toric Threefolds,''
hep-th/9701147.}
\lref\sixman{M. Bershadsky, K. Intriligator, S. Kachru, D. Morrison,
V. Sadov, and C. Vafa, ``Geometric Singularities and Enhanced
Gauge Symmetries,'' Nucl. Phys. {\bf B481} (1996) 215, hep-th/9605200.}
\lref\cdgp{P. Candelas, X. de la Ossa, P. Green, and L. Parkes, 
``A Pair of Calabi-Yau Manifolds as an Exactly Soluble Superconformal
Theory,'' Nucl. Phys. {\bf B359} (1991) 21.}
\lref\ks{S. Kachru and E. Silverstein, ``Singularities, Gauge
Dynamics, and Non-Perturbative Superpotentials in String Theory,''
{Nucl. Phys.} {\bf B482} (1996) 92, hep-th/9608194.}
\lref\threedgauge{O. Aharony, A. Hanany, K. Intriligator, N. Seiberg,
and M.J. Strassler, ``Aspects of N=2 Supersymmetric Gauge Theories
in Three Dimensions,'' hep-th/9703110.}
\lref\fmw{R. Friedman, J. Morgan, and E. Witten, ``Vector Bundles
and F-Theory,'' hep-th/9701162.}
\lref\Misha{
M. Bershadsky, A. Johansen, T. Pantev, and V. Sadov, ``On Four-Dimensional
Compactifications of F-Theory,'' hep-th/9701165.} 
\lref\hw{P. Horava and E. Witten, ``Heterotic and Type I String Dynamics
from Eleven-dimensions,'' Nucl. Phys. {\bf B460} (1996) 506, hep-th/9510209.}
\lref\dketal{J. Distler and S. Kachru, ``(0,2) Landau-Ginzburg Theory,''
Nucl. Phys. {\bf B413} (1994) 213, hep-th/9309110\semi
J. Distler and S. Kachru, ``Duality of (0,2) String Vacua,'' 
Nucl. Phys. {\bf B442} (1995) 64, hep-th/9501111\semi
T. Chiang, J. Distler, and B. Greene, ``Some Features of (0,2) Moduli
Space,'' hep-th/9702030.} 
\lref\aks{O. Aharony, S. Kachru, and E. Silverstein, ``New N=1 Superconformal
Field Theories in Four Dimensions from D-brane Probes,'' Nucl. Phys.
{\bf B488} (1997) 159, hep-th/9610205.}
\lref\eric{E. Sharpe, Princeton preprints  
to appear.}
\lref\edMFphases{E. Witten, ``Phase Transitions in M Theory and
F Theory,'' Nucl. Phys. {\bf B471} (1996) 195, hep-th/9603150.}
\Title{RU-97-27, hep-th/9704185}
{\vbox{\centerline{Chirality Changing Phase Transitions in}
        \vskip4pt\centerline{4d String Vacua}}}
\centerline{Shamit Kachru and Eva Silverstein 
\footnote{$^\star$}
{kachru@physics.rutgers.edu, 
evas@physics.rutgers.edu}}
\bigskip\centerline{Department of Physics and Astronomy}
\centerline{Rutgers University}
\centerline{Piscataway, NJ 08855}

\vskip .3in

We provide evidence that some four-dimensional
$N=1$ string vacua with different numbers of generations
are connected through phase transitions.
The transitions involve going through
a point in moduli space where there is a nontrivial fixed point governing
the low energy field theory.  In an M-theory description, the
examples involve wrapped 5-branes leaving one of the ends of
the world.     

\Date{April 1997} 

\newsec{Introduction}

There has been great progress in the past two years in unifying string
vacua with eight or more supercharges (i.e., the equivalent of 4d N=2
supersymmetry).  For example, extremal transitions between Higgs
and Coulomb phases are believed to connect
all type II compactifications on Calabi-Yau threefolds 
\refs{\oldcand,\andy}, while 
exotic
transitions involving ``tensionless strings'' are believed to connect
vacua with different numbers of tensor multiplets in 6d theories with
minimal supersymmetry \sixd.     

In studying four dimensional vacua with N=1 
supersymmetry, one encounters chiral gauge theories.    
These are very much of interest since the chiral spectrum of quarks
and leptons is one of the basic features of the Standard Model.
In the context of conventional Lagrangian field theory chirality
is preserved by the dynamics, since chiral fermions cannot
pick up masses as long as the gauge symmetry remains unbroken.
It has been widely believed that string vacua with different net numbers of
generations cannot be connected, due to the
absence of a conventional field-theoretic mechanism which can 
provide a mass for 
generations without similarly giving a mass to anti-generations.

In this paper, we propose that in fact vacua of the $E_8 \times E_8$ heterotic
string with 4d N=1 supersymmetry and different generation numbers can be 
connected by phase transitions.  These transitions involve
going through a point in moduli space 
where there is no $\it free$ long-distance description
of the spacetime physics, i.e. there is a nontrivial fixed point.\foot{
The fact that nontrivial fixed points might play an important role in
the explanation of ``conifold'' singularities in heterotic string
theory was suggested some time ago by T. Banks.}
There is already a familiar analogue of this story in six dimensions.  
Passing through nontrivial IR fixed points there
allows one to change the number of tensor multiplets, despite the
absence of a conventional field-theoretic mechanism which can give
these multiplets a mass \sixd.  We argue here 
that in four dimensions, transitions through similar points seem to allow
one to change the net generation number! 
Other interesting 
transitions between N=1 vacua, which do not involve a change
in the net number of generations, have been discussed in 
\refs{\nonchiral,\Misha,\eric,\dketal}.

In \S2, we describe a class of situations in which we believe such a phase
transition occurs.  In \S3, we study in detail a specific example.  We
discuss directions for further exploration in \S4.  

\newsec{Wrapped Fivebranes and Chirality Change}

\subsec{Fibration Picture of the Phase Transition}

Consider the $E_8\times E_8$ heterotic string compactified on
a Calabi-Yau threefold $X$ which is a K3 fibration with a section.  Let 
the base ${\bf P}^1$ have radius $R$.
Above the scale $1/R$ the model reduces to the six-dimensional
theory obtained by compactifying the heterotic string on K3.
We must also specify an appropriate vector bundle 
$\tilde V$ living on the compactification manifold; 
on K3 this amounts to a configuration of Yang-Mills instantons.

This six-dimensional theory develops a nontrivial fixed point
as an instanton shrinks to zero size \sixd.  As explained in \sixd,
this is a multicritical point.  This can be seen by considering
the eleven-dimensional structure of heterotic compactifications \hw.
The vacuum is a compactification on $S^1/{\bf Z}_2\times K3$, with
$E_8$ gauge fields living on the ends of the interval.
The zero-size instanton constitutes a fivebrane on one of the ends.
There are two phases connected by the fixed point theory:
the Higgs phase is obtained by enlarging the instanton,
while a distinct Coulomb-like phase is obtained by 
moving the fivebrane off into the eleven-dimensional bulk.

The fivebrane worldvolume theory contains a tensor multiplet
consisting of a real scalar, an antisymmetric tensor field
with self-dual field strength, and fermions.
The real scalar's VEV gives the distance of the fivebrane
from the end of the world.  
The perturbative heterotic theory has one tensor multiplet,
whose two-form potential combines with that in
the 6d gravity multiplet (which has anti-self-dual
field strength) to form an unconstrained 
antisymmetric tensor field.  
The transition back to the perturbative heterotic 
theory from the phase with extra tensor multiplets
cannot be described in conventional Lagrangian field theory,
since the extra tensors cannot pick up masses without having
any anti-self-dual partners to couple to.  

Of more direct interest in our problem is the fact that
the charged matter content also changes in the transition
from the perturbative heterotic theory to the theory with
the fivebrane in bulk.  For definiteness let us take the 
gauge field configuration of  
the perturbative heterotic compactification on K3 to consist of an 
$SU(2)$ bundle
embedded in one of the 
two $E_8$s, so that the unbroken gauge symmetry is $E_8\times E_7$.
There are then 45 hypermultiplet singlet bundle moduli and
20 ${1\over 2}{\bf 56}s$ of $E_7$.  A single small instanton 
occurs at codimension one in the moduli space.  After
the transition to the theory with the fivebrane in bulk, 
there is one fewer ${1\over 2}{\bf 56}$ in the spectrum.  
This is in accord with the
gravitational anomaly, which requires a loss of 29 hypermultiplets
for each tensor multiplet gained.

Let us now consider the theory in four dimensions, i.e. below
the scale $1/R$.  Again for definiteness let us consider
an $SU(3)$ bundle $V$ embedded in one
$E_8$ on the Calabi-Yau threefold $X$.  This
leads to $E_6$ gauge symmetry in spacetime.  
We will discuss the spectrum in detail
(for a set of examples) in the next section.  The upshot is
that each ${1\over 2}{\bf 56}$ hypermultiplet descends to a set of 
${\bf 27}s$ {\it or} to a  ${\bf\overline{27}}$ in 
the 4d theory.  
There is also a set of ${\bf 27}s$ and one ${\bf\overline{27}}$
which do not descend from matter in the 6d theory, i.e. matter
which is not associated with the generic fiber $(K3,\tilde V)$.  
These are therefore
associated with the singular fibers.

In the four-dimensional theory, 
a zero-size instanton in the generic fiber corresponds
to a fivebrane wrapped around the base ${\bf P}^1$.
As shown in \edsusy,  
it remains consistent with supersymmetry for the fivebrane to
move away from the end of the world, since it is wrapped
on a holomorphic cycle.  In this phase, since one of the
${1\over 2}{\bf 56}s$ has been removed in six dimensions,
the corresponding ${\bf 27}s$ or ${\bf\overline{27}}$ have been removed
in the four-dimensional theory.  We will argue in \S3\ 
that no additional states from singular fibers are produced
in the transition. 
So we find that the net generation number
has shifted.

In \S3.1, we describe how the matter descends from
six to four dimensions on the K3 fibration.  We will see that
the ${1\over 2}{\bf{56}}$s that remain after the 6d phase transition
(and hence their descendant $\bf{27}$s or ${\bf{\overline{27}}}$s) 
can be followed through the transition.  
The matter that gets removed can also be understood physically.
As we will review in \S3.2, the small instanton singularity (like all 
singularities of the worldsheet CFT) can be understood 
explicitly in a linear sigma model formulation of the worldsheet
theory \edphases.  There, the singularity appears as an
extra infinite flat direction in the scalar potential energy.
Most of the vertex operators for spacetime states are not supported down this
throat, and therefore do not ``see'' the singularity.  
The ones that are removed in the transition are supported
down the throat.

The question of matter from the singular
fibers is considered in \S3.3.  We study
a small instanton singularity (which is present in the
generic fiber $(K3,\tilde V)$ as well as on the singular fibers)
in an explicit model.  In this example, we show 
that the small instanton does not
intersect the degenerations of $\tilde V$ in the singular fibers. 
This, along 
with the fact that removing the fivebrane does not
lead to additional singularities, leads us to believe that
the transition does not produce additional matter states
from singular fibers.

One might worry that our proposed transition violates anomaly
matching considerations.\foot{We thank the Rutgers theory
group, in particular T. Banks and N. Seiberg, for enlightening
us on these issues.}  In particular, if we ignore nonrenormalizable
couplings, there appears to be an anomalous $U(1)_R$ symmetry
under which each ${\bf 27}$ and ${\bf \overline{27}}$ transforms 
with charge 2/3.  There is an anomaly (coming from a triangle
diagram with a $U(1)_R$ current and two $E_6$ currents) which
counts the net generation number.  It appears that
the anomaly will not match on the other side of the transition,
since the net generation number has changed.  There is,
however, another chiral multiplet $\Phi$ 
created in the transition--the dimensional
reduction of the 6d tensor multiplet.  So we can 
avoid the problem of matching the anomalies by letting
$\Phi$ transform under the $U(1)_R$.  Then this symmetry
is spontaneously broken on the other side of
the transition, so the issue does not arise.   
Another possibility is that the ${\bf 27}s$ or
${\bf\overline{27}}s$ of interest develop anomalous
dimensions and naively irrelevant nonrenormalizable
terms (which explicitly break
the $U(1)_R$ in string theory)
are actually important for the physics of the transition.

\subsec{Nonperturbative Effects at Lower Scales} 

So far, we have been discussing the four-dimensional theory below
the scale $1/R$ set by the Kaluza-Klein modes on the sphere.
There are in fact lower dynamical scales in the problem, set
by the strength of worldsheet instanton effects.  
In the $E_8 \times E_8$ theory, there are two noncritical strings of
interest in addition to the usual fundamental heterotic string.
These arise from membranes stretched between the fivebrane
and the two ends of the world.  If $\phi$ parametrizes the
$S^1/{\bf Z_2}$, then the tension of these noncritical strings goes like
$\phi$ and $({1\over \alpha^\prime} - \phi)$, respectively.  So   
euclidean worldsheets of these strings wrapping $C$ will generate effects
that go like $e^{-R^2\phi}$ and $e^{-R^{2}({1\over{\alpha^\prime}} - \phi)}$. 
\foot{Some of these ideas were developed in 
discussions with O. Aharony, O. Ganor, and A. Hanany.} 
Similar worldsheet instanton effects, associated with noncritical
strings which arise in F-theory compactifications on Calabi-Yau
fourfolds, have been discussed in \mayr.

The physics of these instanton effects can be classified according to
the splitting type of the vector bundle $V$ over the base $\bf{P}^1$ $C$. 
Actually, in 
general there is another vector bundle $V_2$ embedded in the second
$E_8$.  At the point where the fivebrane is ``absorbed'' 
into the second $E_8$ as an
instanton,  
$V_2$ is also singular over the base curve $C$.
On general grounds, $V$ and $V_2$ split as direct sums of
line bundles at the singularities of interest 
\eqn\vsplit{V \vert_{C} ~=~\sum_{i=1}^{r} {\cal O}(k_i), ~~~V_2 \vert_{C} ~=~
\sum_{j=1}^{s} {\cal O}(l_j)} 
with $\sum_i k_i = \sum_j l_j = 0$ since $c_{1}(V)=c_{1}(V_2)=0$.  
We will now discuss various possibilities for physics of the  
tensionful string worldsheet instantons, sometimes by 
making use of the T-dual $SO(32)$ string
language \refs{\kss,\ks}.    
We have not been exhaustive in our analysis of the possibilities,
and expect interesting new physics to arise for some 
combinations of splitting types of $V$ and $V_2$ that we
have not discussed below.

If $h^{0}(V\vert_{C}\otimes {\cal O}(-1))$ and
$h^{0}(V_2\vert_{C}\otimes {\cal O}(-1))$
are $\it both$ greater than two, then there is no worldsheet
instanton generated potential on the $\phi$ branch.  This follows
from a standard 
counting of worldsheet fermion zero modes \oldjd, which are 
associated with these cohomology groups for the two kinds of 
noncritical strings.  For these splitting types, there are too many
zero modes for a potential to be generated.  Furthermore, in these
cases $\it fundamental$ string worldsheet instantons also cannot
remove the ``small instanton'' point in the classical moduli space, where
the nontrivial fixed point occurs. 
The physics as analyzed at the scale $1/R$ persists in the
full quantum theory.  Such vacua provide examples where the chirality
change can occur while remaining in the moduli space of vacua, i.e. at zero
cost in energy. 

If $h^{0}(V\vert_{C} \otimes {\cal O}(-1))$ or
$h^{0}(V_2\vert_{C} \otimes {\cal O}(-1))$ 
is zero, 
a nonperturbative worldsheet instanton generated superpotential destabilizes
the branch with the fivebrane on the interval.    
One can see this in the $E_8$ theory by considering tensionful string
worldsheet instantons of the first or second $E_8$.  The number of
zero modes on the worldsheet of such a string, for these splitting types,
will be small enough for a nontrivial instanton-generated superpotential.
This is consistent with known results about $SU(2)$ gauge theories
in three dimensions \refs{\ahw,\threedgauge},  
which occur in the T-dual $SO(32)$ description.

In the case where $h^{0}(V\vert_{C}\otimes {\cal O}(-1))=2$ while
$h^{0}(V_2 \vert_{C} \otimes {\cal O}(-1))=0$,  
another interesting 
phenomenon is known to occur, in some examples,  at our small instanton
singularity.  One finds a pole in the Yukawa couplings of the ${\bf 27}s$
or ${\bf\overline{27}}s$ which are supported down the throat
\refs{\cdgp,\silvwitt}.
Before presenting our more detailed evidence for chirality
change in \S3, we will take a detour in \S2.3\ to propose a dynamical
explanation of the poles in Yukawa
couplings in the $E_8 \times E_8$ theory.  

We should note that in the latter two cases of splitting
types we have discussed, unlike the first case, 
the chirality change which seems to be possible at
scales between the scale $\mu$ generated by the
instanton superpotential and $1/R$ is obstructed by quantum effects
below $\mu$.
However, the potential barrier which prevents chirality change in such 
cases is hierarchically smaller than the string scale.
This will become clear in our example in \S2.3.
This indicates that even
in these
cases chirality change can ``almost'' occur, i.e. it is possible at a
small cost in energy.

\subsec{The Pole in the Yukawa Coupling and Noncritical String 
Worldsheet Instantons}

In \kss, the behavior of the $SO(32)$ heterotic string theory on $X$
at a particular small instanton singularity was studied.  The non-perturbative
resolution of the singularity in that case is given by
an $SU(2)$ Yang-Mills theory.
As noted in \ks, the dynamics of the four-dimensional 
fixed point theory obtained by compactifying the $E_8\times E_8$
heterotic string on $X$ at the same type of small instanton 
singularity 
is T-dual, after compactification
on another circle, to the three-dimensional gauge theory obtained
by compactifying the $SO(32)$ heterotic string on 
$X\times S^1$. 
In particular, the two theories are related by $r\to \alpha^\prime/r$
(where $r$ is the radius of the circle)
on the locus where Wilson lines have broken the gauge
symmetry to $SO(16)\times SO(16)$.   

Recently, many aspects of three-dimensional gauge theories have
been explained in \threedgauge.  The example of \kss\ 
involves the $SU(2)$ gauge theory with 
$N_F=2$, which has 4 doublets $d_i,~~ i=1,\dots,4$.  
The superpotential is given by
\eqn\threedsup{W=W_{tree}-Y Pf V+e^{-{R^2\over\alpha^\prime}}Y.}    
Here for large $Y$, $Y=e^\Phi$, where $\Phi$ is the $SU(2)$ Wilson line.
$V_{ij}=\epsilon_{ab}d_i^ad_j^b$ are gauge-invariant combinations
of the doublets.  
A word of explanation is in order concerning the last term
in \threedsup.  In the 3d gauge theory, the last term
is $e^{-1/g_4^2}Y=e^{-1/rg_3^2}Y$, where $g_3$ and $g_4$
are the three and four-dimensional gauge couplings.
The nonperturbative $SU(2)$ gauge coupling in the 
4d $SO(32)$
theory is given by $1/g_4^2=R^2/\alpha^\prime$ \kss.  In applying
T-duality to that case to study our problem, we must
hold $R^2/\alpha^\prime$ constant.  Since we also want to
study the {\it four-dimensional} $E_8\times E_8$ theory, we
are also interested in the limit $r\to 0$, and hence 
$g_3\to\infty$.

In our case, as in \kss, the pole will
be recovered by taking the tree-level superpotential
\eqn\Wtree{W_{tree}=V_{12}{{\bf 10}_1{\bf 10}_1{\bf 1_{-2}}}
+V_{13}^2+V_{14}^2+V_{23}^2+V_{24}^2}
Here we are working on the $SO(16)^2$ locus.  The gauge
symmetry is broken to $SO(16)\times SO(10)\times U(1)$ by 
the $SU(3)$ vector bundle, and each generation consists
of a ${\bf (1,10)_1+(1,1)_{-2}}$.  The first term in
\Wtree\ is proportional to the Yukawa coupling of such a generation.  
Then integrating out $Y$, 
$V_{13},V_{14},V_{23}$, and
$V_{24}$ yields $V_{12}={e^{-{R^2\over\alpha^\prime}}\over V_{34}}$.
Plugging this into \threedsup\ gives
\eqn\Wdyn{W_{dyn}={{{\bf 10}_1{\bf 10}_1{\bf 1_{-2}}
e^{-{R^2\over\alpha^\prime}}\over V_{34}}}}

As we discussed in \S2.2, in the 
original $E_8\times E_8$ setup, $\Phi$ (or more properly
$\phi\equiv \Phi/R^2$) gives the tension of the noncritical string 
which lives on the fivebrane and arises from a membrane
stretched from the fivebrane to the end of the world which has the nontrivial
$SU(3)$ vector bundle.  
Similarly, the tension of the string arising from a membrane
stretched to the other end is ${1\over\alpha^\prime}-\phi$.   
The term $e^{-{R^2\over\alpha^\prime}}e^\Phi$ can be understood
as coming from worldsheet instantons of this string wrapped
around the base ${\bf P}^1$.  
This is the analogue for our problem of the fact, emphasized in
\threedgauge, that the dynamical scale of the 4d gauge theory
appears from ``twisted instantons'' intrinsically in the 3d
theory on a circle.

Notice that in the bosonic potential obtained from \threedsup,
one could take $V=0$ at a cost in energy of 
$M_{string}e^{-{R^2\over\alpha^\prime}}$.  Then $Y$ can be
turned on, leading to a transition at energy scales above
$M_{string}e^{-{R^2\over\alpha^\prime}}$ but well below $1/R$.
In order to be able to freely move the fivebranes, we would
actually
need more than one fivebrane in order to cancel the term
$\vert e^{-{R^2\over\alpha^\prime}}\sum_i Y_i \vert^2$ in 
the potential energy.  Here, $Y_i$ determines the position
of the $ith$ fivebrane.  It is not difficult to find examples
where several fivebranes can move off into the interval giving such
a potential -- the example we discuss in \S3\ is of this
type.

\newsec{An Explicit Example}

In this section, we will explain in some detail the relation
described in \S2.1\ between the six and four-dimensional spectra by
using
standard techniques for the analysis of string compactifications.
We do this in the context of a concrete example.  
The example we discuss below is of the class which
involves smooth chirality change at the scale $1/R$ but in which
there is a potential generated by 
instantons.   
However, we expect the analysis to generalize to a
large class of K3 fibrations and
gauge bundles.

\subsec{Fibration of Charged Matter}

Consider the (0,2) model which is obtained as a deformation of
the (2,2) model on the hypersurface of degree 8 in 
${\bf WP^{4}_{11222}}$.  This is a K3 fibration, with the K3 fiber
realized as a hypersurface of degree 4 in ${\bf P}^3$.
This model was studied in detail in \kss\ using a
gauged linear sigma model on the worldsheet \edphases, and we will
recall from that analysis what we need here.  
Let us denote the projective coordinates on 
${\bf WP^{4}_{11222}}$ by $z_1,\dots,z_5$.  
The hypersurface is given by the vanishing
locus of a degree 8 polynomial $G(z_1,\dots,z_5)$. 
For definiteness, let us take 
\eqn\defeq{G(z_1,\dots,z_5)={z_1^8\over 8}+{z_2^8\over 8}+
{z_3^4\over 4}+{z_4^4\over 4}+{z_5^4\over 4}.}
Setting $z_1=\lambda z_2$ and $y=z_2^2$,
we find the defining equation for the K3 fiber:
\eqn\deffib{\tilde G(y,z_3,z_4,z_5)=
(1+\lambda^8){y^4\over 8}+{z_3^4\over 4}
+{z_4^4\over 4}+{z_5^4\over 4}.}

The bundle is specified by giving polynomials 
$J_i$, $i=1,\dots,5$ of degrees (7,7,6,6,6) respectively.
On the (2,2) locus, the vector bundle becomes the tangent
bundle, so that $J_i={{\partial G}\over{\partial z_i}}$.  
The (0,2) deformation can be obtained by deforming
these polynomials:
\eqn\defJ{\eqalign{
& J_1=p[z_1^7+G_1]\cr
& J_2=p[z_2^7+G_2]\cr
& J_3=p[z_3^3+G_3]\cr
& J_4=p[z_4^3+G_4]\cr
& J_5=p[z_5^3+G_5]\cr}}

Then the vector bundle on the K3 fiber is given by
the polynomials
\eqn\defJfiber{\eqalign{
& F_1={{\lambda J_1+J_2}\over z_2}
\biggr|_{z_1=\lambda z_2, y=z_2^2}=p[(1+\lambda^8)y^3+\tilde G_1]\cr
& F_2=J_3(y,z_3,z_4,z_5)=p[z_3^3+\tilde G_3]\cr
& F_3=J_4(y,z_3,z_4,z_5)=p[z_4^3+\tilde G_4]\cr
& F_4=J_5(y,z_3,z_4,z_5)=p[z_5^3+\tilde G_5]\cr}}

The charged matter (${1\over 2}{\bf 56}s$) in six dimensions
arises partly from monomials of degree 4 in $y,z_3,z_4,z_5$ modulo
the $F_i$.  There are 6 such monomials $A_2$ of the form
$y^2 P_2(z_3,z_4,z_5)$, 7 monomials $A_1$ of the form 
$yP_3(z_3,z_4,z_5)$, and 6 monomials
$A_0$ of the form $P_4(z_3,z_4,z_5)$,
where we have indicated by $P_d(z_i)$ a polynomial of
degree $d$ in $z_i$.  This gives 19 ${1\over 2}{\bf 56}s$.
The index theorem ensures that there are 20; the
additional one will be explained below.
As explained in \vw, we can
read of the transformation properties of these states
under monodromy around points on the base ${\bf P}^1$ above
which the fiber theory is singular.  
Our situation is a generalization of that studied in \vw\ since
in the (0,2) context physical singularities occur when
the {\it vector bundle} degenerates, regardless of the
behavior of the manifold at such points.  
It will simplify our analysis to consider a choice of polynomials such that
none of $F_2,F_3$ or $F_4$ contain a term proportional
to $y^3$.  Then singularities occur 
when $1+\lambda^8=0$.  

Around the points on the base $\bf{P}^1$ with $1+\lambda^{8}=0$, $y$ 
has monodromy
$y\to e^{{{2\pi} i\over 4}}y$.  This leads to monodromies
for the ${1\over 2}{\bf 56}s$ (coming from monomials
$A_2,A_1$, and $A_0$) that we found above.
These transform under the monodromy as
$A_q\to e^{{{2\pi iq}\over 4}}A_q$.  This
means that near a monodromy point $\lambda_0$, 
$A_q\sim (\lambda-\lambda_0)^{q\over 4}$.  
Hence, taking into account all eight singular fibers, we see that
$A_q$ behaves like a polynomial of degree $2q$ in $\lambda$.
Therefore we expect 
$2q+1$ ${\bf 27}s$ in the 4d theory for each 
${1\over 2}{\bf 56}$ corresponding to a monomial $A_q$
in the 6d theory. 

This is indeed what one finds from the analysis of
charged matter in the theory compactified on the full threefold $X$.
In this case we must count monomials of degree 8 
in $z_1,\dots,z_5$ modulo
the $J_i$.  We find $6\cdot 5=30$ monomials of the
form $P_4(z_1,z_2)P_2(z_3,z_4,z_5)$, $7\cdot 3=21$ of
the form $P_2(z_1,z_2)P_3(z_3,z_4,z_5)$, and 6 of
the form $P_4(z_3,z_4,z_5)$.  These correspond to the
expected descendants of $A_2,A_1$, and $A_0$ respectively.
There are additional states in the 4d theory which do
not arise from the ${\bf 56}s$ present in the six-dimensional
theory obtained by compactification on the generic fiber.
Some of these correspond to the 21 monomials of the form 
$P_6(z_1,z_2)P_1(z_3,z_4,z_5)$ and the 5 monomials
of the form $P_8(z_1,z_2)$.  There are also 3
non-deformation-theoretic ${\bf 27}s$.  
  
What about the ${\bf\overline{27}}s$?  One way to understand these
is to use the linear sigma model description reviewed in \kss.
There the coordinates $z_1,\dots,z_5$ are scalar fields transforming
with charges $(1,1,1,1,1)$ and $(1,1,0,0,0)$ under two $U(1)$
gauge symmetries on the worldsheet.  There are, in addition, scalar 
fields $\sigma_1$ and $\sigma_2$ which become part of the (2,2)
gauge multiplets on the (2,2) locus.  On the (2,2) locus,
the gauge multiplets correspond to Kahler moduli.  Each gauge multiplet
yields one $\bf{\overline {27}}$ (whose vertex operator is proportional
to $\sigma$ \silvwitt).  So in our model, there are two
${\bf\overline{27}}s$, one coming from the generic fiber theory
and the other coming from the base ${\bf P}^1$.  
The contribution in the generic fiber gives the requisite
$20th$ ${1\over 2}{\bf 56}$ in the 6d theory.

So we see that each ${1\over 2}{\bf 56}$ present in the
6d theory leads to {\it either} a set of
${\bf 27}s$ {\it or} a ${\bf\overline{27}}$ in the 4d theory.
This suggests, as discussed in \S2, that in a transition
in which a single ${1\over 2}{\bf 56}$ is removed, the
corresponding charged matter in 4d is removed.  
This charged matter being chiral, the phase transition
then changes the net generation number of the 4d theory
(modulo contributions from the singular fibers to be discussed below).

This statement depends on being able to follow the remaining 
${1\over 2}{\bf 56}s$ through the transition in the 6d theory.
We will now discuss the localization of matter both in F-theory
and in the heterotic linear sigma model description.

\subsec{Localization of 6d Matter}
  
The 6d theory is amenable to analysis
using F-theory \VafaF.\foot{F-theory can also be applied to
the study of 4d $N=1$ theories, but it remains
to understand charged matter in such compactifications.}
There the heterotic compactification on K3 with instanton
numbers $12+n$ and $12-n$ in the two $E_8s$ corresponds to
compactification of F-theory on an elliptic Calabi-Yau threefold $K$ with
base $F_n$ \FII.  The manifold $K$ consists of a $T^2$ varying over
$F_n$ (which is itself a ${\bf P}^1$ bundle over ${\bf P}^1$). 
Its defining equation is 
\eqn\defK{y^2=x^3+f(w_1,w_2)x+g(w_1,w_2).}
This describes a $T^2$ at each point $(w_1,w_2)$ on the
base $F_n$.  In general, one has
\eqn\fggen{\eqalign{
& f(w_1,w_2)=\sum_{k=-4}^4 f_{8-nk}(w_2)w_1^{4-k}\cr
& g(w_1,w_2)=\sum_{k=-6}^6 g_{12-nk}(w_2)w_1^{6-k}\cr}} 
where the subscripts on the polynomials $f_{8-nk}$ and $g_{12-nk}$ indicate
their degrees.  The discriminant $\Delta=4f^3+27g^2$ vanishes
at singularities of the $T^2$ fiber.

In our case, we have an $SU(2)$ bundle in one of the $E_8s$, which leaves
an unbroken $E_7 \times E_8$ gauge symmetry in spacetime.  On the F-theory
side, we must therefore induce an $E_7$ singularity 
on the ``top''
${\bf P}^1$ of the $F_n$.  Let us put this singularity at
$w_1=0$.  As explained in \FII, this requires $f\sim w_1^3$ and
$g\sim w_1^5$ near the singularity.  So our polynomials \fggen\ get
restricted to
\eqn\fgEseven{\eqalign{
& f(w_1,w_2)=\sum_{k=-4}^1 f_{8-nk}(w_2)w_1^{4-k}\cr
& g(w_1,w_2)=\sum_{k=-6}^1 g_{12-nk}(w_2)w_1^{6-k}\cr}} 
Near the singularity, the discriminant looks like
\eqn\discEseven{
\Delta\sim w_1^9(f_{8-n}(w_2)+{\cal O}(w_1))}
As explained in \refs{\sixman,\aspg}, the ${1\over 2}{\bf 56}s$ of
$E_7$ are associated with zeroes of the polynomial
$f_{8-n}$. In the case with all instantons in one $E_8$ which
we have been studying, $n=-12$ and there are 20
${1\over 2}{\bf 56}s$.  The $E_8$ symmetry of the
other ``end of the world'' is located at $w_1 = \infty$.

We would now like to study the phase transition, which
involves inducing a small instanton and moving the
resulting fivebrane off the end of the interval.  
As explained in \refs{\FII,\edMFphases}, this is realized in F theory
by the blowing up of a point on the $F_n$.  Following
\S6.1\ of \FII,  we must add a new coordinate $u$, and
mod out by an additional rescaling under which
\eqn\blowup{(x,y,w_1,w_2,u)\to (\lambda^2x,\lambda^3y,
\lambda w_1,\lambda w_2,\lambda^{-1}u).}
This restricts our polynomials to be of the form
\eqn\fgblowup{\eqalign{
& f(w_1,w_2,u)=w_1^3\sum_{l=1}^{8-n}f_{1l}w_2^lu^{l-1}
+{\cal O}(w_1^4)\cr
& g(w_1,w_2)=w_1^5\sum_{l=1}^{12-n}g_{1l}w_2^lu^{l-1}
+{\cal O}(w_1^6)\cr}}
In particular, in order to induce the small instanton
(which makes possible the blow-up), we had to restrict
the polynomials $f$ and $g$ so that $f_{8-n}(w_2)\sim w_2\tilde f_{7-n}$
and $g_{12-n}(w_2)\sim w_2\tilde g_{11-n}$.  The blow-up removes the
point $w_1=w_2=0$, and one is left with $7-n$ 
${1\over 2}{\bf 56}s$ at the zeroes of $\tilde f_{7-n}$.  
Thus the ${1\over 2}{\bf 56}s$ which remain 
can be followed continuously through the transition.  

The heterotic linear sigma model description gives a
simple physical way to characterize the matter fields which get removed
during the transition (at least those coming from
the generic fiber).  The bosonic potential looks like
\eqn\bospot{\eqalign{
&{e_1^2\over 2}\biggl(|z_1|^2+|z_2|^2+2|z_3|^2+2|z_4|^2
+2|z_5|^2-8|p|^2-r_1\biggr)^2\cr
& + {e_2^2\over 2}\biggl(|z_1|^2+|z_2|^2-2|x|^2-r_2\biggr)^2\cr
& +|\sigma_1+\sigma_2|^2\biggl(|z_1|^2+|z_2|^2\biggr)
+|\sigma_1|^2\biggl(64|p|^2+4|z_3|^2+4|z_4|^2+4|z_5|^2\biggr)\cr
& +4|\sigma_2|^2|x|^2+|G|^2+\sum_i|J_i|^2+|J_x|^2\cr}}
Here $p$ and $x$ are scalar fields which decouple in the
generic large-radius phase, $J_x$ is a polynomial
constrained to satisfy $z_1J^1+z_2J^2=2xJ_x$, and
$e_{1,2}$ are the gauge couplings of the two
worldsheet $U(1)$ gauge groups.  As explained in \edphases,
the singularities of the IR conformal field theory have
a beautiful realization in this linear model.  

One type of singularity occurs for special values of
the moduli $r_{1,2}$.  In particular, when $r_1=r_2$, 
there arises a new flat direction in \bospot\ in which 
$\sigma_1$ and $\sigma_2$ can go off to infinity as long
as $\sigma_1+\sigma_2=0$.  On this branch, $z_3,z_4,z_5,p$,
and $x$ are frozen at zero.  Recall that the vertex operator
for the ${\bf\overline{27}}$ which comes from a 
${1\over 2}{\bf 56}$ in six dimensions is given by
$\sigma_1-\sigma_2$.  So the wavefunction for one of
the 4d ${\bf\overline{27}}s$ is supported along the
``throat'' which produces the singularity.  On the other
hand, the ${\bf 27}s$ coming from 
${1\over 2}{\bf 56}s$ in the 6d theory arise from monomials involving
$z_3, z_4$, and $z_5$.  These are frozen at zero way out
on the throat where $\sigma_1$ and $\sigma_2$ diverge.  
These can therefore not be affected by the transition involving
the fivebrane.         

A similar distinction between the types of states arises
at singularities obtained by specializing the parameters
in the $J_i$ to induce a singularity on the generic fiber.
Then for example $z_3$ and $p$ might diverge along a new
branch in \bospot, forcing $\sigma_1=0$.  So only (a subset
of) the ${1\over 2}{\bf 56}s$ which lead to ${\bf 27}s$ can
be removed in the phase transition.

\subsec{The Singular Fibers}

We would now like to give the promised explicit verification, 
in an example, that the small instanton does not intersect
the singularity associated with the singular fibers.
In our model \defJfiber, we have singularities at 
$\lambda=\lambda_0$ such that
$1+\lambda_0^8=0$, since there the $F_i$ can all vanish
without $y$ vanishing.  These $\lambda_0$ are the points
on the base ${\bf P}^1$ over which there are singular fibers.
The singularity in the fiber theory occurs at $z_3=z_4=z_5=0$.
Let us take for simplicity 
$\tilde G_1=yz_3^2$,
$\tilde G_3=e^{-{i\pi\over 2}}z_3z_4^2-y^2z_3$, and
$\tilde G_4=-e^{-{i\pi\over 2}}z_4z_3^2$ with 
$\tilde G_5$ vanishing.  Then there
are singularities of the bundle (which amount to
some collection of small instantons) at $y=z_5=0;z_3=e^{{{i\pi k}\over 4}}z_4$
(for $k=1,3,5,7$).  These small instanton singularities do not
intersect the singularity $z_3=z_4=z_5=0$ of the singular fiber.
As explained in \S3.2, the removal of a small instanton
from a fiber K3 does not seem to introduce
any additional singularities.  For instance, in the F-theory description
which renders the whole analysis geometrical, the removal
of the fivebrane corresponds to blowing up a point and
does not introduce new singularities.
Hence, it appears very implausible that 
removing the small instanton fivebranes could change the
contributions from the singular fibers.    

Because of the subtleties involving the singular fibers, 
it would be very interesting to develop techniques for describing
directly the vector bundle which remains on the end of the world after the
transition.    Perhaps this can be done using appropriate generalizations of
\refs{\fmw,\Misha}\ to chiral models.

\newsec{Discussion}

We presented evidence here that in a specific class of M-theory 
vacua -- compactifications on $S^1/{\bf Z}_2$ times a
K3 fibration-- there
seem to be chirality-changing phase transitions.  It is of significant
interest for the vacuum degeneracy problem to understand
the extent to which this phenomenon occurs in more general
vacua.  

In particular, heterotic compactifications typically develop
pointlike singularities at some codimension in moduli space.
The nonperturbative physics at such singularities has yet
to be determined.  
It was shown long ago that naively, (2,2) models can
be connected at such singularities \oldcand.  The Euler
character $\chi$ of the manifold changes in the transition.
In context of type II string theories, $\chi$ gives
the difference between the numbers of hypermultiplets and
vector multiplets in the resulting 4d $N=2$ theory.
The transitions of \oldcand\ correspond, as explained
in \andy, to transitions between Higgs and Coulomb phases
in 4d $N=2$ gauge theory.

In the context of the heterotic string, $\chi$ gives
the net generation number for (2,2) models, and the transitions of
\oldcand\ cannot occur within conventional (weakly coupled) Lagrangian
field theory.  
In the example discussed in this paper, the occurence of a nontrivial
fixed point theory enables branches with different numbers of
generations to meet.
If the nonperturbative physics which resolves
the ``conifold'' singularities of \oldcand\ in 
heterotic string compactifications
also  
involves nontrivial interacting fixed points,  
one can 
reasonably speculate that branches of the moduli space with different
numbers of generations may be meeting there as well.\foot{Note that
unlike the case studied here, where the transition is mediated by emission 
of a wrapped 
M theory fivebrane, it is difficult to see how one would understand 
directly what
is on the ``other side'' of possible transitions in the pointlike conifolds.} 

It is attractive to conjecture that there is a web of heterotic
(0,2) and (2,2) vacua, much like the web of N=2, d=4 theories, connected
by transitions through nontrivial fixed points (such as those found in
\aks).
In fact, given that there are already known transitions which connect 
(0,2) heterotic vacua on
topologically distinct manifolds (but at fixed generation number)
purely $\it within$ perturbative string theory \dketal, it is 
very interesting to continue the study
of nonperturbative transitions between (0,2) models 
with different generation numbers on a 
$\it fixed$ Calabi-Yau manifold.

\smallskip

\centerline{\bf Acknowledgements}
\smallskip

We would like to thank O. Aharony, P. Aspinwall, T. Banks, M. Douglas, 
O. Ganor, A. Hanany, J. Louis, N. Seiberg, and S. Shenker 
for helpful discussions.  This work was supported in part by
DOE DE-FG02-96ER40559.

\listrefs
\end